%% file: short_paper.tex
\documentclass[sigconf, screen, nonacm, timestamp=false]{acmart}

\usepackage{enumitem}

\AtBeginDocument{%
  \providecommand\BibTeX{{%
    \normalfont B\kern-0.5em{\scshape i\kern-0.25em b}\kern-0.8em\TeX}}}

\acmConference[EASE 2025]{The 29th International Conference on Evaluation and Assessment in Software Engineering}{17–20 June, 2025}{Istanbul, Türkiye}
\copyrightyear{2025}
\acmYear{2025}
\setcopyright{acmlicensed}
\acmDOI{10.1145/XXXXXXX.XXXXXXX}
\acmISBN{978-1-4503-XXXX-X/18/06}

\begin{document}

\title{Analyzing the Usage of Donation Platforms for PyPI Libraries}

\author{Alexandros Tsakpinis}
\affiliation{%
  \institution{fortiss GmbH}
  \city{Munich}
  \country{Germany}}
\email{tsakpinis@fortiss.org}

\author{Alexander Pretschner}
\affiliation{%
  \institution{Technical University of Munich}
  \city{Munich}
  \country{Germany}}
\email{alexander.pretschner@tum.de}

\input{sections/0_Abstract}
\input{sections/css-concepts}
\keywords{OSS Funding, OSS Libraries, Repository Mining}

\maketitle

\input{sections/1_Introduction}
\input{sections/2_Related_Work}
\input{sections/3_Research_Method}
\input{sections/4_Results}
\input{sections/5_Discussion}
\input{sections/6_Threats_to_Validity}
\input{sections/7_Conclusion_and_Future_Work}

\bibliographystyle{ACM-Reference-Format}
\bibliography{short_paper}

\appendix

\end{document}

%% file: sections/0_Abstract.tex
\begin{abstract}
    Software systems extensively rely on open source software (OSS) libraries, which offer numerous benefits but also pose significant risks. These risks arise when vulnerabilities or attacks emerge, and the OSS community fails to address them promptly due to inactivity or lack of resources.
    Recent research highlights the strong connection between OSS maintenance activities and financial support. To support the sustainability of the OSS ecosystem, it is crucial for library maintainers to register on donation platforms and link these profiles on the library's project page accordingly. This allows end users and industry initiatives to provide financial support, ensuring maintainers have access to funding streams. However, a comprehensive investigation on the actual usage of donation platforms in OSS ecosystems is currently missing.
    This descriptive study analyzes the usage of the most common donation platforms in the PyPI ecosystem. For every available PyPI library, we retrieve its assigned URLs, direct dependencies, and, when available, the owner type and additional donation platform links from its GitHub repository. Using the PageRank algorithm, we analyze the ecosystem for different subsets of libraries looking at both the library and dependency chain perspective.
    Our study provides several empirical insights regarding the adoption of donation platforms within the PyPI ecosystem. We observe that donation platform links are largely omitted from PyPI project pages, with a strong preference for listing such links exclusively on GitHub repositories. Additionally, GitHub Sponsors emerges as the dominant donation platform, though a notable portion of listed links on PyPI are outdated, highlighting the need for automated link verification. Our findings also reveal significant variations in donation platform adoption across individual libraries and dependency chains. While the analysis of individual PyPI libraries exhibit relatively low adoption rates, libraries used as direct and transitive dependencies show a much higher usage of donation platforms. This widespread adoption of donation platforms among dependencies is a positive sign for developers using PyPI libraries, as these libraries actively seek financial support to sustain ongoing maintenance.
\end{abstract}

%% file: sections/css-concepts.tex
\ccsdesc[500]{Software and its engineering~Maintaining software}

%% file: sections/1_Introduction.tex
\newpage
\section{Introduction}
Over time, Open Source Software (OSS) has become a cornerstone of the product life cycle, driven by commercial, engineering, and quality reasons \cite{ebert2008open}. Today, OSS components make up between 80 and 90\% of the code in commercial products, underscoring their extensive adoption across the software industry \cite{pittenger2016open, oss2022}. A critical part of OSS are software libraries — often referred to as packages — which play a pivotal role in modern software development. Libraries allow developers to reuse established, well-tested code for specific functionalities, reducing the need for building everything from scratch~\cite{bauer2012structured}. Once integrated into a project, a library becomes a dependency \cite{cox2019surviving}, often forming a dependency chain where each dependency relies on other dependencies to function properly. However, OSS dependencies are not without risks. They can be affected by vulnerabilities (e.g., Log4j\footnote{\url{https://logging.apache.org/log4j/2.x/}}) or supply chain attacks (e.g., XZ Utils\footnote{\url{https://openssf.org/blog/2024/03/30/xz-backdoor-cve-2024-3094/}}, colors\footnote{\url{https://snyk.io/de/blog/open-source-npm-packages-colors-faker/}}) posing significant challenges to the software industry \cite{decan2018impact}. When such issues arise, the library's community typically responds promptly by releasing a fixed version~\cite{rahkema2022swiftdependencychecker}.

However, support for a library can sometimes be terminated or suspended, leading to the significant challenge of no maintenance activities being available \cite{bauer2012structured, raemaekers2011exploring}. Even worse, issues with a single unmaintained library can cascade through the interconnected OSS ecosystem, affecting thousands of others through direct and transitive dependencies \cite{kula2014visualizing, tsakpinis2023analyzing}. While research from academia and industry emphasize the importance of funding OSS to sustain maintenance activities, maintainers must first leverage funding infrastructure, such as donation platforms, to be eligible for financial support \cite{tidelift2024, medappa2023sponsorship}. Popular donation platforms include GitHub Sponsors, Open Collective, Patreon, Ko-fi, Buy Me a Coffee, Liberapay, and PayPal. These platforms serve as critical funding infrastructure for open-source maintainers, allowing them to register and potentially receive direct financial support from end users and industry initiatives. Companies like Microsoft\footnote{\url{https://github.com/microsoft/foss-fund}\label{microsoft_funding}}, Stripe\footnote{\url{https://resources.github.com/open-source/why-stripe-sponsors-open-source/}\label{stripe_funding}}, and Indeed\footnote{\url{https://engineering.indeedblog.com/blog/2019/07/foss-fund-six-months-in/}\label{indeed_funding}} run funding programs that select OSS maintainers for financial support, often requiring them to be registered on such donation platforms to qualify. To enhance visibility and improve their chances of receiving support, PyPI maintainers can add donation platform links to the "Project Links" section on their PyPI project page, as illustrated by an example library\footnote{\url{https://pypi.org/project/devtools/}\label{link_example}}. Yet, while this option exists, little is known about how OSS maintainers actually leverage these donation platforms within the PyPI ecosystem. To address this, the ecosystem should be analyzed from two perspectives: first, by analyzing libraries individually without considering their dependencies. Second, by including dependency chains in the analysis, which is crucial due to the cascading effects of dependencies. The following research questions (RQs) are designed to gather evidence and provide these insights:

\vspace{1mm}
\textbf{RQ1:} What is the distribution of the most commonly used donation platforms across the PyPI ecosystem?

\textbf{RQ2:} What is the ratio of individual PyPI libraries which link to donation platforms?
 
\textbf{RQ3:} What is the ratio of PyPI libraries in dependency chains which link to donation platforms?
\vspace{1mm}

To address these research questions, this study focuses on the PyPI ecosystem due to its widespread use in modern software development and the strong integration of libraries into its development practices \cite{decan2016topology, abdalkareem2020impact, Octoverse2024}. We gather assigned URLs from the project page on PyPI, along with available details on the repository owner's type (individual or organization) and any donation platform links provided on GitHub. We focus on GitHub as the code management system, given its continued popularity for hosting code of OSS libraries — particularly those used in PyPI \cite{eghbal2020working, tsakpinis2024analyzing}. Since libraries are highly interconnected, analyzing them individually provides limited insights. Instead, we also analyze their dependency chains. To achieve this, we collect direct dependencies for each library, constructing a dependency graph enriched with donation platform information. Recognizing that the importance of libraries within their ecosystem can vary widely, we calculate each library's importance using its PageRank score derived from the graph \cite{tsakpinis2024analyzing, mujahid2021toward}. This dependency graph forms the foundation for analyzing the usage of donation platforms from both a library and dependency chain perspective using subsets of libraries with varying importance.

Our key contributions are several empirical insights into different aspects of the PyPI ecosystem. First, we find that only 3.1\% of identified donation platform links are actively listed on PyPI, while the remaining 96.9\% appear exclusively on GitHub repositories. This suggests a strong tendency among PyPI maintainers to omit donation platform links from their PyPI project pages. Furthermore, the distribution of donation platform links on PyPI and GitHub is heavily skewed towards
GitHub Sponsors, which accounts for 81.3\% of all identified links. Notably, among the GitHub Sponsors URLs linked on PyPI, 28.6\% were outdated, underscoring the need for an automated URL checker offering an opt-in option to alert maintainers when links become obsolete. Our analysis also highlights variations in the adoption of donation platforms across individual libraries and dependency chains. Among individual PyPI libraries, up to 14.3\% utilize donation platforms — 9.6\% managed by organizations and 4.7\% by individuals. The adoption rate increases significantly within dependency chains: up to 32.0\% of libraries rely on donation platforms when analyzing libraries including their direct and transitive dependencies. This increases to 72.8\% for direct dependencies and 76.4\% for transitive dependencies when being analyzed separately. This disparity suggests that libraries, which are not used as dependencies, mostly lag in donation platform adoption, presenting an opportunity for improvement. However, the high adoption rates in direct and transitive dependencies provide a positive signal for developers using PyPI libraries, as most dependencies, which are integrated automatically without direct control, actively seek financial support to sustain ongoing maintenance.

%% file: sections/2_Related_Work.tex
\section{Related Work}
The following section presents related work to emphasize the importance of the broader topic of funding OSS ecosystems and to highlight how this study differs from and contributes to existing research. It covers studies about OSS ecosystems from diverse perspectives, and studies specifically addressing OSS funding.

Many studies explore package dependency networks for various ecosystems, focusing on security issues like insertion of malicious code \cite{guo2023empirical}, vulnerabilities \cite{dusing2022analyzing, alfadel2023empirical} and the evolution of ecosystem robustness \cite{hafner2021node}. Research also addresses package versioning, update strategies \cite{javan2023dependency}, configuration challenges~\cite{peng2023less}, dependency conflicts~\cite{wang2020watchman}, code smells \cite{cao2022towards}, and errors \cite{mukherjee2021fixing}. Additional studies investigate the characteristics of widely used NPM packages~\cite{mujahid2023characteristics}, the factors influencing sustainable OSS Python projects~\cite{valiev2018ecosystem} and the use of trivial packages in PyPI and NPM \cite{abdalkareem2020impact}. While these studies analyze individual libraries in OSS ecosystems, none have explored the usage of donation platforms, which this study aims to address.

While many studies focus on specific ecosystem aspects, others examine broader structural features, such as dependencies~\cite{bommarito2019empirical, decan2016topology}, with particular attention to transitive dependencies \cite{decan2019empirical, kikas2017structure, tsakpinis2024analyzing}. However, the usage of donation platforms in dependency chains remains unexplored, creating a gap that this study intends to fill.

In the area of OSS funding, studies have examined topics such as the impact of public funding on open-source development \cite{osborne2024toolkit}, the effects of sponsorship funding \cite{medappa2023sponsorship}, and the state of open source maintainers \cite{tidelift2024}. The latter study, surveying over 400 independent OSS maintainers, reveals that about 25\% of the respondents receive support from donation platforms. However, no research has utilized actual ecosystem data to analyze the usage of donation platforms. By applying our data-driven approach, this study provides a more objective and comprehensive view on the situation, complementing and enhancing the insights from existing survey-based findings.

%% file: sections/3_Research_Method.tex
\section{Research Method}
The following section explains our methodology for this descriptive study, focusing on how the data collection and analysis were conducted, with the primary goal of exploring the usage of donation platforms among PyPI libraries.

\subsection{Data Collection}
\label{sec:data_collection}
To evaluate the usage of donation platforms within the PyPI ecosystem, we collect library information from different sources with various Python scripts. This requires first a comprehensive list of available libraries, which is provided via an endpoint\footnote{\url{https://pypi.org/simple/}\label{all_endpoint_pypi}}. For detailed information about each library, including assigned URLs and dependency data, another endpoint\footnote{\url{https://pypi.org/pypi/<package\_name>/json}\label{detailed_endpoint_pypi}} is used. We collect all direct and optional dependencies for the latest library version to create a dependency graph with unlimited transitive depth. For assigned GitHub repository URLs, the GitHub GraphQL API\footnote{\url{https://api.github.com/repos/<repository\_owner>/<repository\_name>}\label{repo_endpoint_github}} is used to identify the repository owner type as either an individual or an organization, and collect additional donation platform URLs linked to the repository. We also check for each GitHub repository URL whether the owner is enrolled in GitHub Sponsors, in case the link is missing or outdated on PyPI. The final output is a JSON file mapping library names to their direct dependencies, associated URLs, and, where available, information about GitHub Sponsors profiles, additional donation platform links and repository owner types.

\subsection{Data Analysis}
In our data analysis, we employ standard terminology such as nodes, dependency chains, and dependency graph. Each library functions as a node within this graph, with dependency chains connecting each node to its direct dependencies. Collectively, this network forms a dependency graph that represents the ecosystem. Data from Section \ref{sec:data_collection} form the basis for several analyses, providing insights from two angles. First, viewing each library as a standalone node — without factoring in dependency chains — offers a library focused perspective on the ecosystem. Analyzing nodes individually provides a simple view of each library. Second, analyzing the data from a dependency chain perspective highlights the paths originating from each node in the graph and unveils a deeper understanding of the ecosystem's structure, revealing insights that are not visible when considering libraries in isolation. For the library perspective, we calculate the following first two metrics based on the research questions mentioned before, while the third one is computed for the dependency chain perspective:

\begin{itemize}[leftmargin=1.5em]
    \item Distribution of the most commonly used donation platforms
    \item Distribution of individual libraries linking to donation platforms
    \item Ratio of libraries in dependency chains linking to donation platforms
\end{itemize}

The second and third metrics are calculated for different subsets of the dependency graph. To define these subsets, we apply the PageRank algorithm, which assigns an importance value to each node in the graph based on its incoming and outgoing dependencies, similar to \cite{tsakpinis2024analyzing, mujahid2021toward}. Categorizing libraries by their importance allows for a more nuanced understanding and enhances the analysis by differentiating between libraries at various levels of importance.

%% file: sections/4_Results.tex
\section{Results}

Data collection began on February 3, 2025, and lasted about a week due to GitHub's API rate limits, resulting in a dataset of 605,504 PyPI libraries. While parsing the dependencies of each library, we excluded libraries not found in the initial ecosystem's library list\textsuperscript{\ref{all_endpoint_pypi}}. Thus, 1.1\% of libraries from a node perspective and 1.3\% from a dependency perspective were removed. Additionally, 3.1\% of libraries were excluded due to errors encountered while querying the detailed ecosystem endpoint\textsuperscript{\ref{detailed_endpoint_pypi}}, primarily resulting from 404 status codes. These errors occur if the package name retrieved from the endpoint of the comprehensive library list\textsuperscript{\ref{all_endpoint_pypi}} could not be found at the detailed endpoint\textsuperscript{\ref{detailed_endpoint_pypi}}. The final dataset consists of 586,866 PyPI libraries that collectively rely on 1,433,490 dependencies, originating from 79,714 distinct libraries.

The analysis concentrates on libraries with increasing importance, defined by their PageRank score. Libraries with higher scores have greater influence within the dependency graph and are more frequently used. In Figures \ref{fig:results_node_perspective} and \ref{fig:percentage_nodes_having_funding_provider}, the x-axis defines the subsets used for calculations. For example, $x=10$ refers to the top 10\% of libraries ranked by the PageRank algorithm, while $x=100$ includes all libraries in the dataset, spanning both high and low importance. Results are often expressed as value ranges relating to the different subsets. Also, if not stated otherwise, the terms funding and donation are used interchangeably in the following paragraphs.

\subsection{RQ1+RQ2: Library Perspective}
\label{sec:node_perspective}

\begin{figure}[h]
    \centering
    \includegraphics[width=0.48\textwidth, trim=0 2mm 0 2mm, clip]{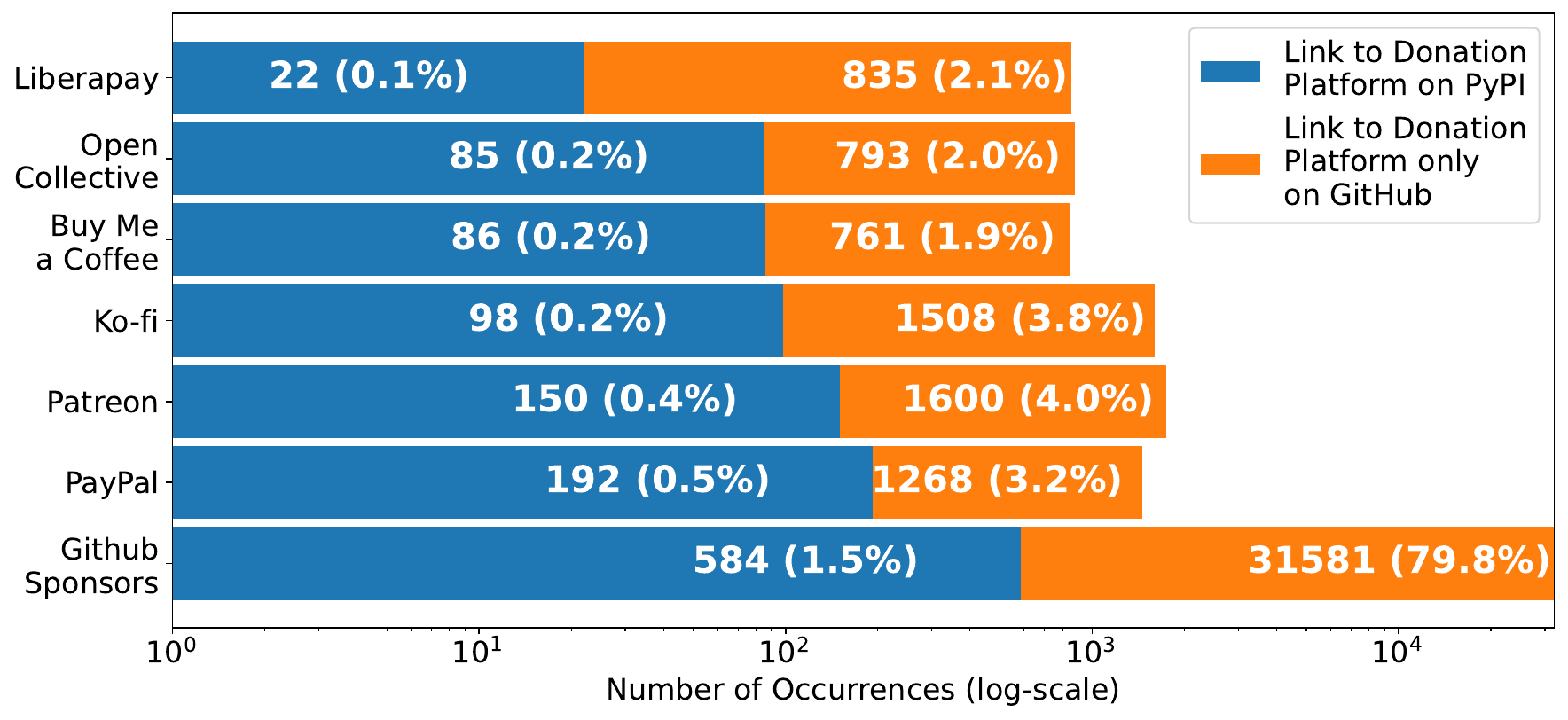}
    \caption{Distribution of used donation platforms (log-scale)}
    \label{fig:results_funding_providers}
\end{figure}

\begin{figure}[h]
    \centering
    \includegraphics[width=0.48\textwidth, trim=0 2mm 0 2mm, clip]{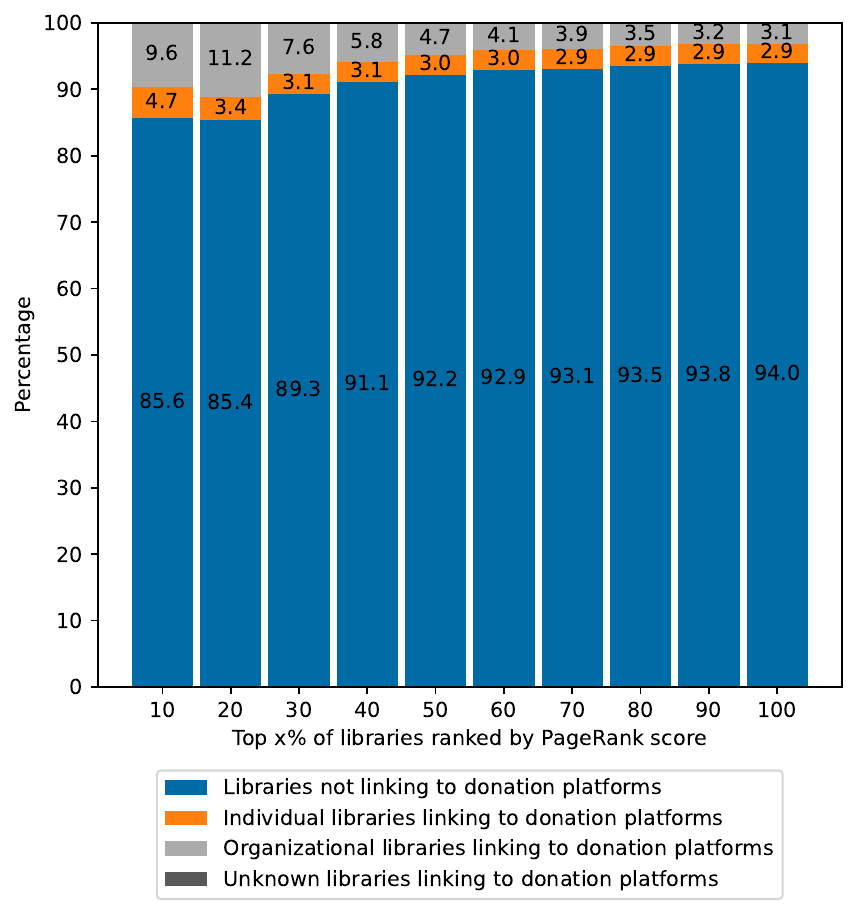}
    \caption{Distribution of individual PyPI libraries linking to donation platforms}
    \label{fig:results_node_perspective}
\end{figure}

Figures \ref{fig:results_funding_providers} and \ref{fig:results_node_perspective} present the results, focusing on the characteristics of individual nodes in the dependency graph without considering their dependency chains. Specifically, the first analysis explores the distribution of donation platforms across all links assigned to PyPI libraries. The second analysis examines the distribution of libraries utilizing donation platforms, distinguishing between libraries developed by individuals and those developed by organizations.

Figure \ref{fig:results_funding_providers} illustrates the distribution of donation platforms utilized within the PyPI ecosystem. The data is derived by collecting all available links associated with each PyPI library. These links are gathered either from the PyPI project page or if available from its GitHub repository. Finally, we convert the links to lowercase and count the occurrences of each domain name. Domains appearing at least 20 times were manually categorized to determine if they qualify as donation platforms. This threshold was validated by comparing the identified platforms with those listed under the "Donations" category on \url{https://www.oss.fund/}, a curated directory of donation platforms. Throughout the rest of the paper, the term "donation platform" will refer to the platforms shown in Figure~\ref{fig:results_funding_providers}.

Overall, among the 586,866 PyPI libraries, 39,563 links were classified as donation platforms. Only 3.1\% (1,217, blue) of these were explicitly added by maintainers on PyPI, while the vast majority — 96.9\% (38,346, orange) — were exclusively found on their GitHub repositories. Links, which are exclusively shown on GitHub repositories, have not been listed on the PyPI project page by the library maintainer but were automatically identified through an associated GitHub repository URL found on PyPI. This imbalance indicates that maintainers prefer linking to donation platforms on GitHub rather than PyPI, supporting the logarithmic scale used in Figure \ref{fig:results_funding_providers}.

Among the links found only on GitHub (orange), GitHub Sponsors is the most common, accounting for 79.8 percentage points (pp) (31,581). It is followed by Patreon (4.0 pp, 1,600), Ko-fi (3.8 pp, 1,508), PayPal (3.2 pp, 1,268), Liberapay (2.1 pp, 835), Open Collective (2.0 pp, 793), and Buy Me a Coffee (1.9 pp, 761). Surprisingly, many maintainers do not take the extra step of adding their donation platform URL to their PyPI project page, even though they have already included their GitHub repository URL. If a GitHub repository URL is provided, PyPI could offer an opt-in option to detect and link any donation platform associated with the repository.

Among the links actively assigned on PyPI (blue), GitHub Sponsors leads with 1.5 pp (584), followed by PayPal (0.5 pp, 192), Patreon (0.4 pp, 150), Ko-fi (0.2 pp, 98), Buy Me a Coffee (0.2 pp, 86), Open Collective (0.2 pp, 85) and Liberapay (0.1 pp, 22). Notably, 28.6\% (167) of the 584 GitHub Sponsors links were outdated, emphasizing the need for an automated URL checker offering an opt-in option to notify maintainers about changes — an issue previously highlighted in related work \cite{tsakpinis2024analyzing}. This result is not visible in any Figure.

By combining the results for GitHub Sponsors URLs found on PyPI and GitHub, we observe a strong preference for GitHub Sponsors, which accounts for 81.3\% of the total links ($584+31,581=32,165$). When considering only links identified on GitHub (38,346, orange), GitHub Sponsors represents 82.4\% (31,581), whereas it accounts for 48.0\% (584) of the links exclusively found on PyPI (1,217,  blue). This preference may stem from GitHub Sponsors' ease of use and the reliance of various industrial funding programs on its availability such as Microsoft’s FOSS Fund\textsuperscript{\ref{microsoft_funding}}, Stripe\textsuperscript{\ref{stripe_funding}}, and Indeed\textsuperscript{\ref{indeed_funding}}.

After identifying the most frequently used donation platforms, Figure \ref{fig:results_node_perspective} shows the breakdown of libraries that use these platforms compared to those that do not. Among the libraries that do include donation platforms, they are further categorized into those developed by individuals and those created by organizations. Overall, only 6.0\% of libraries include links to donation platforms — 2.9\% by individual developers and 3.1\% by organizations.

Focusing on more important libraries by shifting from $x=100$ to $x=10$ in Figure \ref{fig:results_node_perspective}, the percentage rises to 14.3\% for the top 10\% of libraries. This increase, however, differs by developer type: individually developed libraries grow to 4.7\%, while organizationally developed libraries increase to 9.6\%. This is unexpected, as individuals are assumed to have a greater need for funding due to working alone. Although the actual reasons remain unclear, one explanation may be that organizations can distribute tasks more effectively, enabling them to select someone specifically to manage donation platform registrations. Additional experiments, not shown in any Figure, further confirm that the positive trend for individually and organizationally developed libraries continues when focusing on the top 1\% of the most important libraries. Given the critical role these libraries play in the PyPI ecosystem, they are expected to actively seek financial support to ensure ongoing maintenance, which should be reflected in the results. Results reveal that only 24.2\% of libraries are linked to donation platforms, with 7.8 pp attributed to libraries developed by individuals and 16.4 pp to those developed by organizations. This finding is both surprising and concerning, as it suggests that out of the 587 most important libraries, only about 142 link to any donation platform.

Another interesting observation is that most libraries linking to a donation platform on PyPI also provide a valid GitHub repository URL. This is reflected in the almost complete absence of entries in the last category ("\textit{Unknown libraries linking to donation platforms}" $\leq 0.1\%$). This finding is promising for related research \cite{tsakpinis2024analyzing}, as it suggests that libraries concerned with their funding situation also enable automated monitoring of their maintenance activities on GitHub through frameworks like the OpenSSF Scorecard\footnote{\url{https://github.com/ossf/scorecard}\label{openSSF}}.

\subsection{RQ3: Dependency Chain Perspective}
\label{sec:dependency_chain_perspective}

\begin{figure}[h]
    \includegraphics[width=0.48\textwidth, trim=0 2mm 0 2mm, clip]{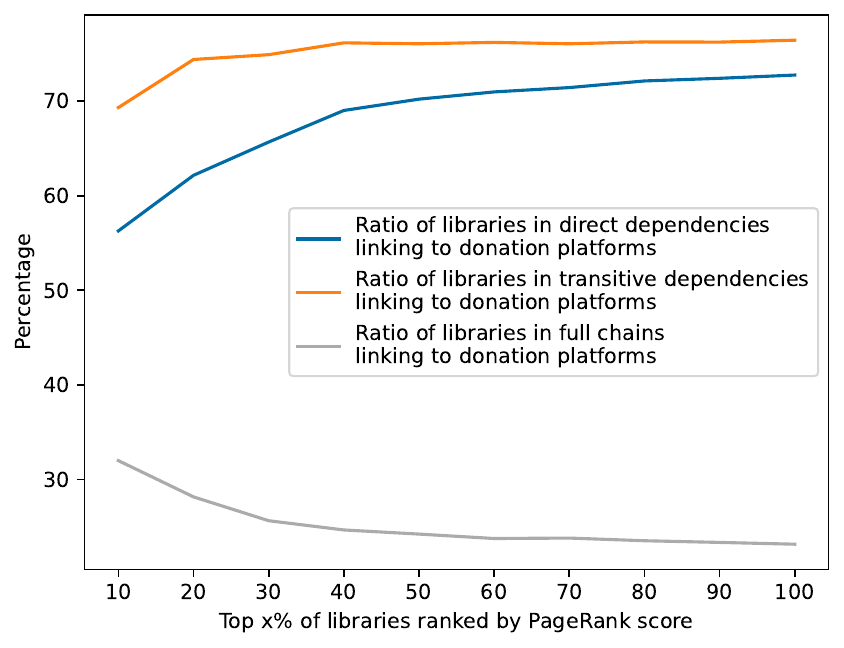}
    \caption{Ratio of libraries in dependency chains linking to donation platforms}
    \label{fig:percentage_nodes_having_funding_provider}
\end{figure}

Figure \ref{fig:percentage_nodes_having_funding_provider} presents the results of analyzing the libraries and their dependency chains from three perspectives, focusing on the ratio of libraries that are linked to a donation platform. Specifically, the analysis considers: (1) each library in the PyPI ecosystem along with its direct and transitive dependencies, referred to as \textit{full chain}; (2) only the direct dependencies of each PyPI library; and (3) only the transitive dependencies. The distinction between direct and transitive dependencies is essential as direct dependencies are explicitly chosen and actively managed, whereas transitive dependencies are automatically included and beyond direct control. The results for each perspective are aggregated across all libraries to provide a comprehensive view of the entire ecosystem.

Figure \ref{fig:percentage_nodes_having_funding_provider} illustrates that as the importance of libraries increases, between 23.2\% and 32.0\% of libraries in full chains are linked to donation platforms. Within direct dependencies, this proportion rises to between 56.3\% and 72.8\%, while in transitive dependencies, it increases further to between 69.3\% and 76.4\%. Despite the upward trend for more important libraries in full chains, the overall percentages remain low, suggesting that only up to a third of libraries in full dependency chains are eligible for donations on these platforms. However, the ratios within direct and transitive dependencies are consistently higher than those for full chains. This reveals that libraries, which are not used as dependencies, often lack donation platforms, while their direct and transitive dependencies are more likely to have links to donation platforms assigned. Therefore, the missing donation platforms in full chains could be directly addressed by the library maintainers themselves, as they control the metadata of their libraries uploaded to the ecosystem - such as links to donations platforms or repositories. The high ratio of libraries in direct and transitive dependencies may serve as a positive signal for developers using PyPI libraries as most dependencies, which are integrated automatically without direct control, are making efforts to secure ongoing maintenance through the usage of donation platforms.

%% file: sections/5_Discussion.tex
\section{Discussion}
We analyzed the usage of donation platforms across the PyPI ecosystem from a node and dependency chain perspective, experimenting with different subsets of libraries varying their level of importance. We focus on libraries with increasing importance, as they occupy central roles in their ecosystems and exert substantial influence.

From a library perspective, the adoption of donation platforms among PyPI libraries is alarmingly low, even for the most important ones. Of the top 587 libraries, representing the top 0.1\%, only 24.2\% have a donation platform linked on PyPI or GitHub. Across the entire ecosystem, this figure drops to just 6.0\%. Regardless of the importance of the libraries in the analyzed subset, individually developed libraries are less frequently linked to donation platforms compared to those developed by organizations, which may benefit from shared resources and collaborative efforts in setting up such platforms. While many library maintainers include donation platform links in their GitHub repositories, they rarely add them to their PyPI project pages, with 96.9\% of donation platform links being exclusively listed on their GitHub repositories. Moreover, GitHub Sponsors strongly dominates the landscape, representing 81.3\% of all donation platform links identified. Competitors may react to this imbalance by identifying areas for improvement and implementing strategies to attract more users to their platforms. Additionally, a closer examination of the GitHub Sponsors URLs linked on PyPI reveals that 28.6\% of them are outdated. In summary, library maintainers should prioritize linking their libraries to donation platforms — not just on their GitHub repository, but also on their PyPI project page in order to be eligible for funding from end users and industry initiatives. This may increase visibility and eventually improve their chances of receiving funding. Also, PyPI could offer an opt-in option to automatically link to any donation platform when available for a linked GitHub repository, and periodically verifying that links remain up-to-date.

However, when evaluating the usage of donation platforms, the dependency chain perspective is crucial as libraries rarely exist in isolation without having dependencies, further complementing the view on the PyPI ecosystem. The ratios of libraries linking to donation platforms in direct and transitive dependencies show promising results with values reaching 72.8\% and 76.4\%. Interestingly, the majority of missing donation platforms exist in libraries, which are not used as dependencies, with values of up to 32.0\%. This is a gap that package maintainers could address directly themselves. The results about direct and transitive dependencies may serve as a positive signal for developers using PyPI libraries as most dependencies, which are integrated automatically without direct control, are making efforts to secure ongoing maintenance through the usage of donation platforms.

%% file: sections/6_Threats_to_Validity.tex
\section{Threats to Validity}
The threats to validity can be categorized into four key aspects \cite{runeson2009guidelines}:

\textbf{Construct validity:}
For the primary study objective of analyzing the usage of donation platforms among OSS libraries on PyPI, a minor threat to construct validity arises when selecting a specific set of donation platforms. This risk is minimized by deriving the selection solely from objective patterns identified in the dataset, such as counting the frequency of each domain name and manually categorizing those that qualify as donation platform. Another potential threat stems from our approach to resolving library dependencies. We ignore version information for any dependency, always opting for the latest version. While this may not fully capture the dynamics of the broader ecosystem, we argue that any resulting discrepancies in the calculated PageRank score of a library are minimal.

\textbf{Internal validity:} 
Threats may stem from data‑handling errors (e.g., mis‑tagged URLs, missed dependencies). We mitigated them by rerunning the data collection and spot‑checking random samples.

\textbf{External validity:} 
We concentrate on PyPI as the ecosystem and GitHub as the code management system (CMS). While the findings might not apply universally to other ecosystems and CMSs, the research methodology applied can be adapted to any publicly accessible ecosystem where libraries can incorporate links, like those to donation platforms and code repositories, within their metadata. We did not consider alternative CMSs, as previous research showed that PyPI libraries strongly prefer linking to GitHub repositories over alternatives like GitLab or BitBucket \cite{tsakpinis2024analyzing}.

\textbf{Reliability:} 
Although we gathered all data from publicly accessible sources, which theoretically supports reproducibility, a threat to reliability arises from the dynamic nature of the PyPI ecosystem. The data represent a specific snapshot from the collection day, which conflicts with reproducibility since PyPI does not offer any time travel functionality. To facilitate full reproducibility, the collected data and analysis code are made available on figshare \cite{tsakpinis_pretschner_2025}. 
Another risk comes from the frequent introduction or modification of libraries, which could lead to gaps or outdated information in our dataset. To mitigate this, the study should be repeated regularly.

%% file: sections/7_Conclusion_and_Future_Work.tex
\section{Conclusion and Future Work}
The paper analyzes how maintainers of PyPI libraries utilize donation platforms on PyPI and GitHub. These platforms play a crucial role in providing infrastructure for financial support, which is vital for the sustainability of OSS projects, as a lack of funding opportunities can lead to reduced maintenance activities, ultimately increasing security risks. We analyzed the PyPI ecosystem from a library and dependency chain perspective, offering a comprehensive overview of how donation platforms are utilized. From a library perspective, we discovered that donation platform links are mostly missing on PyPI project pages, with a clear tendency to list them on GitHub repositories instead. GitHub Sponsors stands out as the primary donation platform across PyPI and GitHub. From a dependency chain perspective, we discovered that a significant proportion of libraries link to donation platforms — up to 72.8\% in direct dependencies and 76.4\% in transitive ones — demonstrating that most dependencies within the PyPI ecosystem are eligible to receive funding from end users and industry initiatives.

Future work should investigate why authors and maintainers choose to link, or not link, their libraries to donation platforms. We also aim to understand why GitHub Sponsors is the preferred choice among these donation platforms and why donation platform links are primarily shared on GitHub repositories rather than directly on PyPI. Additionally, we plan to regularly repeat this study to track ecosystem changes regarding the use of the donation platforms.